\documentclass[12pt]{article}
\usepackage{epsfig}
\usepackage{a4,amsmath}
\begin{document}
\begin{center}
\begin{Large}
{\bf{Concerning the generalized
 Lorentz symmetry and the  generalization of the Dirac 
 equation}} \footnote{The research is supported by DFG grant
No.\,436\,RUS.}\\
\end{Large}   
\vskip 5mm
\begin{large}
      G.Yu. Bogoslovsky $^{a,\,}$\footnote{{\it{E-mail address:}} 
      bogoslov@theory.sinp.msu.ru},
    \,H.F. Goenner $^{b,\,}$\footnote{{\it{E-mail address:}}
      goenner@theorie.physik.uni-goettingen.de} \\
\end{large}
\vskip 3mm
 {\footnotesize{$^{a}$ {\it{Skobeltsyn Institute of Nuclear Physics, Moscow State University\\
                            119992 Moscow, Russia}}\\
                $^{b}$ {\it{Institute for Theoretical Physics, University of G\"ottingen,
                            37081 G\"ottingen, Germany}}}}\\

\end{center}
\vskip 7mm
\hrule
\vskip 3mm
\begin{small}
\noindent
{\bf{Abstract}}
\vskip 2mm
   The work is devoted to the generalization of the Dirac equation for a flat 
locally anisotropic, i.e., Finslerian space-time. At first we reproduce the corresponding metric and a 
group of the generalized Lorentz transformations, which has the meaning of the relativistic symmetry 
group of such event space. Next, proceeding from the requirement of the generalized Lorentz invariance 
we find a generalized Dirac equation in its explicit form. An exact solution of the nonlinear 
generalized Dirac equation is also presented.\\
\end{small}
\vskip 1mm
{\footnotesize{ 
\noindent
{\sl{PACS\,:}}\, 03.30.+p; 03.65.Pm; 11.30.Cp; 11.30.Qc; 02.20.Sv; 02.40.Hw \phantom{{\it{Keywords}}\,: }Spontaneous symmetry breaking
                 
}}                 
\vskip 3mm
\hrule
\vskip 2cm

In spite of the impressive successes of the unified gauge theory of strong, weak and 
electromagnetic interactions, known as 
the Standard Model, one cannot a priori rule out the 
possibility that Lorentz symmetry underlying the theory is an approximate symmetry of nature. 
This implies  that at the energies already attainable today empirical evidence may be obtained in 
favour of violation of Lorentz symmetry. At the same time it is obvious that such effects 
might manifest themselves only as strongly suppressed effects of Planck-scale physics.

Theoretical speculations about a possible violation of Lorentz symmetry continue for more than 
forty years and they are briefly outlined in [1]. Nevertheless we 
note here that, along with the spontaneous breaking [2], one of the 
first and, as it appeared subsequently, fruitful ideas relating to a possible violation of 
Lorentz symmetry was the idea [3] according to which the 
metric of event space differs from Minkowski metric and the physically equivalent inertial
reference frames are linked by some transformations which differ from Lorentz ones. In 
[4] such transformations were called generalized Lorentz 
transformations. Note also that the idea about the existence of generalized Lorentz 
transformations was suggested in connection with the situation in the physics of ultra-high 
energy cosmic rays, namely, with the absence of the Greisen-Zatsepin-Kuz'min effect (the 
so-called GZK cutoff) predicted [5, 6] on the 
basis of conventional relativistic theory. The absence of the GZK cutoff has yet not been 
explained convincingly and still remains the main empirical fact which indirectly speaks in 
favour of violation of Lorentz symmetry.

Interest in the problem of violation of Lorentz and CPT symmetries has revived in recent years 
[7] in connection with the construction of a phenomenological theory 
reffered to as the Standard-Model Extension [8].

In the present work, which is in essence devoted to the same problem, we proceed from the 
assumption [9] that phase transitions with breaking of
gauge symmetries should be accompanied by phase transitions in the geometric structure of 
space-time.

Our study is based on the fact [10] 
that the Lorentz symmetry is not the only possible
realization of the relativistic symmetry. Another admissible realization of the 
relativistic symmetry is obtained with the aid of nonunimodular matrices 
belonging to a group of the  generalized Lorentz transformations. 
In contrast to the conventional Lorentz transformations, the  generalized ones 
conformally modify Minkowski metric but leave invariant the corresponding 
Finslerian metric which describes a flat locally anisotropic space-time. Thus, from the
formal point of view the locally anisotropic space-time appeares as the necessary
consequence of the existence of a group of the generalized Lorentz 
transformations. As for the physical nature of the anisotropy, there are some
reasons to suppose that a fermion-antifermion condensate, which may arise 
[11] (instead of elementary Higgs condensate) in the 
spontaneous breaking of initial gauge symmetries, turns out to be anisotropic and its 
anisotropy determines the local anisotropy of event space. Obviously, verification of 
this hypothesis is far from being trivial.
Therefore the opening investigations in this direction, as presented here, are aimed at 
the most fundamental problem, namely, at the generalization of the Dirac equation for 
the locally anisotropic space-time.

Consider the metric [4] of a flat locally anisotropic 
space-time
\begin{equation}\label{1}
ds^2=\left[\frac{(dx_0-\boldsymbol\nu d\boldsymbol x)^2}{dx_0^2-d\boldsymbol x^{\,2}}\right]^r
(dx_0^2-d\boldsymbol x^{\,2})\,.
\end{equation}
Being not a quadratic form but a homogeneous function of the coordinate
differentials of degree two, this metric falls within the category of
Finsler metrics [12]. It depends on two constant parameters $r$ and 
$\boldsymbol\nu$, in which case the unit vector $\boldsymbol\nu$ indicates a preferred direction in
3D space while $r$ determines the magnitude of space anisotropy, characterizing the degree of
deviation of the metric (1) from the Minkowski metric. Thus the anisotropic 
event space (1) is the generalization of the isotropic Minkowski space of 
conventional special relativity theory.

The 3-parameter noncompact group of the generalized Lorentz transformations, which leave
the metric (1) invariant, appears as
\begin{equation}\label{2}
x'\,^i=D(\boldsymbol v,\boldsymbol\nu )\,R^i_j(\boldsymbol v,\boldsymbol\nu )\,L^j_k(\boldsymbol v)\,
x^k\,,
\end{equation}
where $\boldsymbol v$ denotes the velocities of moving (primed) inertial reference frames; the matrices 
$\,L^j_k(\boldsymbol v)\,$ represent the ordinary Lorentz boosts; the matrices 
$\,R^i_j(\boldsymbol v,\boldsymbol\nu )\,$ represent additional rotations of the spatial axes of 
the moving frames around the vectors $\,[\boldsymbol v\,\boldsymbol\nu ]\,$ through the angles
\begin{equation}\nonumber
\varphi=\arccos\left\{ 1-\frac{(1-\sqrt{1-\boldsymbol v^{\,2}/c^2})
[\boldsymbol v\boldsymbol\nu ]^2}{
(1-\boldsymbol v\boldsymbol\nu /c)\boldsymbol v^{\,2}}\right\}
\end{equation} 
of relativistic aberration of $\boldsymbol\nu\,;$ and the diagonal matrices
\begin{equation}\nonumber
D(\boldsymbol v,\boldsymbol\nu )=\left(\frac{1-\boldsymbol v\boldsymbol\nu /c}
{\sqrt{1-\boldsymbol v^{\,2}/c^2}}
\right)^rI\,,
\end{equation}
the additional dilatational transformations of the event coordinates. The structure of the
transformations (2) ensures the fact that in spite of a new geometry 
of event space the 3-velocity space remains Lobachevski space.
 
With the inclusion of the 1-parameter group of rotations about $\,\boldsymbol\nu\,$ and of the
4-parameter group of translations the inhomogeneous isometry group of
the Finslerian event space (1) turns out to have eight parameters. 
If the third spatial axis is chosen along $\,\boldsymbol\nu\,,$ then its generators can be written as 
\begin{equation}\nonumber
\begin{array}{rcl}
X_1&=&-(x^1p_0+x^0p_1)-(x^1p_3-x^3p_1)\,,\\
X_2&=&-(x^2p_0+x^0p_2)+(x^3p_2-x^2p_3)\,,\\
X_3&=&-rx^ip_i-(x^3p_0+x^0p_3)\,,\\
R_3&=&x^2p_1-x^1p_2\,; \qquad \qquad \qquad \qquad \qquad p_i=\partial /\partial
x^i\,.\\
\end{array}
\end{equation} 
These generators satisfy the commutation relations
\begin{equation}\nonumber
\begin{array}{llll}
  [X_1X_2]=0\,, & [R_3X_3]=0\,, && \\
  \left[X_3X_1\right]=X_1 \,, & [R_3X_1]=X_2 \,, && \\
  \left[X_3X_2\right]=X_2\,, & [R_3X_2]=-X_1\,; & & \\
  \left[p_i p_j\right]=0\,; &&& \\
  \left[X_1p_0\right]=p_1\,,& [X_2p_0]=p_2\,, & [X_3p_0]=rp_0+p_3\,, &
  [R_3p_0]=0\,, \\
  \left[X_1p_1\right]=p_0+p_3\,,& [X_2p_1]=0\,, & [X_3p_1]=rp_1\,, &
  [R_3p_1]=p_2\,, \\
  \left[X_1p_2\right]=0\,, & [X_2p_2]=p_0+p_3\,, & [X_3p_2]=rp_2\,, &
  [R_3p_2]=-p_1\,, \\
  \left[X_1p_3\right]=-p_1\,, & [X_2p_3]=-p_2\,, & [X_3p_3]=rp_3+p_0\,, &
  [R_3p_3]=0\,. \\
\end{array}
\end{equation}  
As one can see the 8-parameter inhomogeneous isometry group of the space-time 
(1) is a subgroup of the 11-parameter extended Poincar\'e (similitude) 
group [13] whereas the homogeneous one is isomorphic 
to the corresponding 4-parameter subgroup 
(with the generators $X_1\,, X_2\,, X_3\!\!\mid_{r=0}\,, R_3$) of the homogeneous Lorentz group.
It is shown in [14] that the 6-parameter homogeneous Lorentz group 
has no 5-parameter subgroup while the 4-parameter subgroup is unique (up to isomorphisms).
Thus, the transition from Minkowski space to the Finslerian event space (1) 
implies a minimum of symmetry-breaking of the Lorentz symmetry. However the relativistic
symmetry is maintained in the form of the generalized Lorentz symmetry.

Because of nonunimodularity of the matrices ${\cal L}^i_k=D R^i_j L^j_k$
representing the generalized Lorentz transformations (2), the transformational
properties of some geometric entities turn out to be changed as compared with
conventional special relativity theory. For instance, a 4-volume element $dx^0d^3x$ is
no longer invariant but is a scalar density of weight $-1\,,$ i.e., it transforms as
follows: $\,dx'{}^0d^3x'=J^{-1}dx^0d^3x\,,$ where $J$ is the Jacobian,
$\,J=|\partial x^k/\partial x'{}^j|=|{{\cal L}^{-1}}^k_j|=D^{-4}\,.$ Similarly,
matrices $\,{\eta}_i{}_k\,$ and $\,{\eta}^i{}^k\,$ having the identical forms
$\,{\eta}_i{}_k=diag\,(1,-1,-1,-1)\,$ and $\,{\eta}^i{}^k=diag\,(1,-1,-1,-1)\,$ in all
frames of reference related by the transformations (2) 
are no longer invariant tensors
but are, respectively, a covariant tensor density of weight $-1/2$ and a contravariant
tensor density of weight $1/2.$ This statement signifies that
$\,{{\eta}'}_i{}_k=J^{-1/2}{{\cal L}^{-1}}^l_i{{\cal L}^{-1}}^m_k{\eta}_l{}_m=  
{\eta}_i{}_k\,$ and $\,{{\eta}'}^i{}^k=J^{1/2}{\cal L}^i_l{\cal L}^k_m{\eta}^l{}^m=
{\eta}^i{}^k\,.$ Then it is clear that $\,{\eta}_i{}_k{\eta}^k{}^l={\delta}^l_i\,$
is a unit tensor. Later on we shall be using $\,{\eta}_i{}_k\,$ and $\,{\eta}^i{}^k\,$
to lower and raise indices. The process, however, will be accompanied by a change in
weight. We shall be also in need of an entity $\,{\nu}^i\,$ which indicates a preferred
direction in 4D space-time and whose components have the same values, 
$\,\left\{\,{\nu}^0=1\,, 
\,\boldsymbol\nu\,\right\}\,,$ in all frames of reference 
related by the transformations (2). It is easy to verify that 
$\,{\nu}'{}^i=J^{(1+r)/(4r)}{\cal L}^i_k{\nu}^k={\nu}^i\,,$ i.e., that $\,{\nu}^i\,$ is
a contravariant vector density of weight $\,(1+r)/(4r)\,,$ in which case $\,{\nu}_i{\nu}^i=0\,$ is
an invariant equation. 

Using the $\,{\nu}^i\,$ and $\,{\eta}_i{}_k\,$ one can represent the metric 
(1) as an explicit invariant of the transformations 
(2)\,: $\,ds^2=\left[({\nu}_idx^i)^2/dx_jdx^j\right]^rdx_kdx^k\,.$ 
With the aid of this expression we arrive at the relativistically invariant action 
for a free particle in the flat anisotropic space. The action and its variation appear as 
\begin{equation}\nonumber 
S=-m\int\limits_a^bds\,, \quad \delta S=-\int\limits_a^bp_i\,d\,\delta x^i\,.
\end{equation}
Hereafter we put $\,c = \hbar = 1\,.$ The principle of least action under the condition 
$\,(\delta x^i)\,|_a=(\delta x^i)\,|_b=0\,$ leads to $\,p_i=const,$ i.e., to rectilinear inertial 
motion. And if one varies the coordinates of point $b$ under the condition $\,p_i=const\,,$ then 
$\,p_i=-\partial S/\partial x^i\,,$ i.e., $\,p_i\,$ is a canonical 4-momentum. Since $\,{\eta}^i{}^k\,$ 
is a contravariant tensor density of weight $1/2$ and since $\,p_k\,$ is a covariant vector it is clear 
that the 4-momentum $\,p^i=(p^0,\,\boldsymbol{p})={\eta}^i{}^kp_k\,$ is transformed as a contravariant 
vector density of $1/2,$ i.e.,
\begin{equation}\label{3}
{p'}^{\,i}=J^{1/2}{\cal L}^i_k\,p^k=D^{-1} R^i_l L^l_k\,p^k\,.
\end{equation}
Thus we have arrived at the generalized Lorentz transformations for 4-momenta. Note that the 
dilatational transformations of $\,p^i\,$ are inverse to those of $\,x^i\,$ 
(~cf.\,(2) )\,. In an explicit form
\begin{equation}\label{4} 
p^i=m\left(\frac{\nu_ldx^l}{\sqrt{dx_jdx^j}}\right)^r\left((1-r)\frac{dx^i}{\sqrt{dx_adx^a}}+
r\frac{\nu^i\sqrt{dx_bdx^b}}{\nu_ndx^n}\right)\,. 
\end{equation}
Since the direction of $\,p^i\,$ is not aligned with the direction of $\,dx^i\,$ we introduce 
(apart from $\,p^i\,$) the so-called kinematic 4-momentum $\,k^i\,$ which has the same 
transformational properties as $\,p^i\,.$
\begin{equation}\label{5} 
 k^i=m\,(1-r)\frac{dx^i}{\sqrt{dx_ndx^n}}\left(\frac{\nu_ldx^l}{\sqrt{dx_jdx^j}}\right)^r\,.
\end{equation}
Taking into account the equation  $\,{\nu}_i{\nu}^i=0\,$ we obtain the following relations
\begin{equation}\label{6} 
p^i=k^i+\frac{r\,k_lk^l}{(1-r)k_n\nu^n}\,\nu^i\,, \qquad 
k^i=p^i-\frac{r\,p_lp^l}{(1+r)p_n\nu^n}\,\nu^i\,.
\end{equation}
As for the 3-velocity of a particle, it is determined by the formula 
$\,\boldsymbol v=\boldsymbol k/k^0\,$. The components of canonical 4-momenta satisfy the mass 
shell equation
\begin{equation}\label{7}
\left[({\nu}_ip^i)^2/p_jp^j\right]^{-r}p_ap^a=m^2(1-r)^{(1-r)} (1+r)^{(1+r)} \,.
\end{equation}
This equation is an invariant of the transformations (3). The same 
mass shell but in a space of kinematic 4-momenta is described by the equation
\begin{equation}\label{8}
\left[({\nu}_ik^i)^2/k_jk^j\right]^{-r}k_ak^a=m^2(1-r)^2 \,.
\end{equation}
The last two equations lead us to the important conclusion, namely, that the motion of free massless
particles in anisotropic space is similar to their motion in isotropic
space, i.e., massless particles do not perceive the space anisotropy whereas the motion of massive 
particles  is analogous to that of quasiparticles in a crystalline medium.

According to (8), the mass shell for $\,r\ne 0\,$ is a deformed 
two-sheeted hyperboloid inscribed into a light cone. In order to show how its deformation changes 
with the magnitude $\,r\,$ of space anisotropy it is reasonable to proceed from the relations 
(5) which determine four-dimensional coordinates of points belonging to 
the upper sheet of the deformed hyperboloid as explicit functions of 3-velocities 
$\,\boldsymbol v =d\boldsymbol x/dx^0\,.$  
The results of calculations, presented in Fig.\,1 clearly 
demonstrate the fact that, if $\,r\to 1\,,$ the mass shell in a space of kinematic 4-momenta converges 
(nonuniformly) to a light cone.
As for the canonical momenta $\,p^i\,,$ there is nonuniform convergence\,: $\,p^i\to k^i+m{\nu}^i\,,$ 
where $\,k_ik^i=0\,.$ Physically this means that the effective inertial mass of a particle present 
in anisotropic (Finslerian) space depends on the magnitude $\,r\,$ of a constant anisotropy field 
and disappears at all if $\,r\,$ reaches the value equal to unity.

Thus, with a view to generalizing the Dirac Lagrangian for the Finslerian space-time 
(1), we have arrived 
at the following guiding principle\,: a generalized Lagrangian, in the limit $\,r=1\,,$ must be 
reducible (up to a 4-divergence) to the standard massless Dirac Lagrangian.

Starting to generalize the Dirac Lagrangian, first consider the standard massless one 
$\,(i/2)\left(\bar\psi{\gamma}^n{\partial}_n\psi -{\partial}_n
\bar\psi{\gamma}^n\psi\right)\,.$ Since 
massless particles do not perceive the space anisotropy, the Lagrangian considered 
need not be modified and it can be used as the kinetic term of a massive generalized Dirac Lagrangian. 
Since under the generalized Lorentz transformations (2) the 4-volume 
$\,dx^{0}d^{3}x\,$ behaves as a scalar density of weight $\,-1\,$ and the action must remain invariant, 
it follows that the kinetic term (\,just as the entire Lagrangian\,) must be a scalar density of 
weight $\,1\,.$ This condition is fulfilled in the case where the generalized Lorentz transformations (2) of the coordinates are accompanied by the following transformations of 
the fields $\,\psi \,$ and $\,\bar{\psi}\,:$
\begin{equation}\label{9}
\psi' (x') =D^{-3/2} S_RS_L \psi(x)=J^{3/8}S\psi(x)\,,
\end{equation}
\begin{equation}\label{10}
\bar{\psi}'(x') = \bar{\psi}(x)J^{3/8}S^{-1}\,,
\end{equation}
where the matrices $\,S=S_RS_L\,$ satisfy the standard condition 
$\,S^{-1}\gamma^iS = \Lambda^i_k \gamma^k\,,$ in which case $\,\Lambda^i_k=R^i_j L^j_k\,;$ the 
matrices $\,S_L\,$ and $\,S_R\,$  represent, respectively, the Lorenz boosts and 
additional rotations of bispinors. In an explicit form
\begin{equation}\label{11}
S_L=\sqrt{\frac{1\!+\!\sqrt{1\!-\!{\boldsymbol v}^2}}{2\sqrt{1\!-\!{\boldsymbol v}^2}}}\,I -
      \sqrt{\frac{1\!-\!\sqrt{1\!-\!{\boldsymbol v}^2}}{2\sqrt{1\!-\!{\boldsymbol v}^2}}}\,\,
      \frac{{\gamma}^0\boldsymbol v\boldsymbol{\gamma}}{|\boldsymbol v|}\,,
\end{equation}
\begin{equation}\label{12}
S_R=\sqrt{1\!-\!\frac{(1\!-\!\sqrt{1\!-\!{\boldsymbol v}^2})[\boldsymbol v\boldsymbol\nu]^2}
      {2(1\!-\!\boldsymbol v\boldsymbol\nu )\,{\boldsymbol v}^2}}\,I +
      i\sqrt{\frac{1\!-\!\sqrt{1\!-\!{\boldsymbol v}^2}}{2(1\!-\!\boldsymbol v\boldsymbol\nu )}}
      \,
      \frac{[\boldsymbol v\boldsymbol\nu]}{|\boldsymbol v|}\,\boldsymbol\Sigma \,,
\end{equation}
where $\,\boldsymbol v\,$ denotes the velocities of moving (primed)  reference frames, 
$\,{\gamma}^n\,$ are the Dirac matrices, 
$\,\boldsymbol\Sigma =diag\,(\,\boldsymbol\sigma\,, \boldsymbol\sigma\,)\,$ 
and $\,\boldsymbol \sigma\,$ are the Pauli matrices. Thus in the flat Finslerian space-time 
(1) the fields $\,\psi \,$ and $\,\bar{\psi}\,$ are, according to 
(9) and (10), bispinor density fields of weight $\,3/8\,.$

In order to generalize the massive term $\,-m\bar{\psi}\psi \,$ of the Dirac Lagrangian
we remind that a generalized massive term, like the kinetic one, must be  a scalar density of 
weight $1\,.$ It can be verified that for the bispinor density fields\,: $\,\bar{\psi}\psi\,$ 
is a scalar density of weight $\,3/4\,,$ $\,\,\bar{\psi}{\gamma}^n\psi\,$ is a contravariant vector 
density of weight $1\,,$ $\,[({\nu_n \bar{\psi}\gamma^n\psi}/{\bar{\psi}\psi})^2]^{r/2}
\bar{\psi}\psi\,$ is a scalar density of weight $\,1\,$ and 
$\,[({\nu_n \bar{\psi}\gamma^n\psi}/{\bar{\psi}\psi})^2]^{-3r/2}
\bar{\psi}\psi\,$ is a scalar, in which case the latter Finslerian form generalizes the scalar 
bilinear form $\,\bar{\psi}\psi\,$ of conventional theory.

Now we are able to write down a Lagrangian for
the bispinor density fields representing such a generalization of the standard
Dirac Lagrangian that the corresponding field equations turn out to be
invariant under the group of generalized Lorentz transformations. It appears as
\begin{equation}\nonumber
{\cal L}= \frac{i}{2}\left(\bar\psi{\gamma}^n{\partial}_n\psi -{\partial}_n
\bar\psi{\gamma}^n\psi\right) - m\!\left[\!\left(
\frac{\nu_n \bar{\psi} \gamma^n \psi}{\bar{\psi}\psi}
\right)^{\!\!\!2} \right]^{\!r/2}\!\!\bar{\psi}\psi  \,.
\end{equation}
This Lagrangian leads to the following generalized Dirac equations\,:
\begin{equation}\label{13}
i \gamma^a{\partial}_a\psi  
\!-\!m\!\!\left[\!\left( 
\frac{\nu_n j^n}{\bar{\psi}\psi}
\right)^{\!\!\!2}  \right]^{\!\frac{r}{2}}\!\!\!\left\{\!\!(1\!-\!r)I\!+\!r
 \!\left(\frac{\bar{\psi}\psi}{\nu_n j^n}\right)\!\nu_a
\gamma^a  \!\right\}\!\psi = 0\,,
\end{equation}
\begin{equation}\label{14}
i {\partial}_a\bar{\psi}\gamma^a  
\!+\!m\!\!\left[\!\left( 
\frac{\nu_n j^n}{\bar{\psi}\psi}
\right)^{\!\!\!2}  \right]^{\!\frac{r}{2}}\!\!\bar{\psi}\!\left\{\!\!(1\!-\!r)I\!+\!r
 \!\left(\frac{\bar{\psi}\psi}{\nu_n j^n}\right)\!\nu_a
\gamma^a  \!\right\} = 0\,,
\end{equation}
where $\,j^n=\bar{\psi}{\gamma}^n\psi \,.$ 
The operation\,: $\,\bar{\psi}$ (13) + (14) 
$\psi\,$ provides the equation $\,\partial_nj^n=0\,.$ Thus $\,j^n\,$ is a preserved current. 
And at last, owing to the
operation\,: $\,\bar{\psi}$ (13) -- (14) 
$\psi \,,$ we conclude that $\,{\cal L}=0\,$ on the solutions of eqs. (13) and (14).

Due to translational invariance, the generalized Dirac equations (13),  
(14) must admit solutions in the form of plane waves 
$\,\psi (x) = u(p)\exp{(-ip_ax^a)}\,.$ This means that the amplitude $\,u(p)\,$ must satisfy the 
equations\,:
\begin{equation}\label{15}
p_a\!\left[\gamma^a\!-\!\left(\frac{\bar u\gamma^au}{\bar uu}\right)\!\!\left\{\!(1\!-\!r)I\!+\! 
r\!\left(\frac{\bar uu}{\nu_l\bar u\gamma^lu}\right)\nu_n\gamma^n\right\}\right]\!u  = 0\,,
\end{equation}
\begin{equation}\label{16}
\bar up_a\!\left[\gamma^a \!-\! \left(\frac{\bar u\gamma^au}{\bar uu}\right)\!\!\left\{\!(1\!-\!r)I\!+
\! r\!\left(\frac{\bar uu}{\nu_l\bar u\gamma^lu}\right)\nu_n\gamma^n\right\}\right]  = 0\,,
\end{equation}
\begin{equation}\label{17}
p_a\bar u\gamma^au = m\left[\!\left(
\frac{\nu_l \bar u\gamma^lu}{\bar uu}
\right)^{\!\!\!2} \right]^{\!r/2}\!\!\bar uu\,.
\end{equation}
Eqs. (15)--(17) lead to the invariant dispersion 
relation
\begin{equation}\label{18}
\sqrt{p_ap^a/(1-r^2)}=\pm\,m\left[(1+r)({\nu}_ip^i)^2/((1-r)p_jp^j)\right]^{r/2}\,,
\end{equation}
where the sign$\,+$ corresponds to positive frequency states whereas the sign$\,-$ corresponds 
to negative ones. It is worth mentioning that the mass shell equation (7) 
can be obtained from (18). In order to find the planewave solutions in a 
general form, i.\! e., at arbitrary momentum \!$p^a$ 
we, for a start, confine ourselves to the rest frame and 
try the following ansatz\,:
\begin{equation}\nonumber
{\psi}_{+}(x)=\sqrt{2m(1\!-\!r)}\left(\begin{array}{cc} \varphi \\ 0 \end{array} 
\right)e^{-i(mx^0-mr\boldsymbol\nu\boldsymbol x)}\quad ,
\end{equation}
\begin{equation}\nonumber
{\psi}_{-}(x)=\sqrt{2m(1\!-\!r)}\left(\begin{array}{cc} 0 \\ \chi \end{array} 
\right)e^{i(mx^0-mr\boldsymbol\nu\boldsymbol x)}\quad ,
\end{equation}
where $\,\sqrt{2m(1\!-\!r)}\,$ is a normalizing multiplier and $\,\varphi\,, \,\chi\,$ are
arbitrary 3-spinors normalized by means of $\,{\varphi}^\dagger \varphi =1\,, \,
{\chi}^\dagger \chi =1\,.$ It is easy to verify that the corresponding positive and negative 
frequency bispinor density amplitudes satisfy Eqs. (15)--(17).
Note once more that these solutions  are found in the rest frame, 
in which $\,p^a=\{m\,, rm\boldsymbol\nu\}\,,$ whereas kinematic 4-momentum
$\,k^a=\{m(1-r)\,, \boldsymbol 0\}\,$ and, respectively, $\,\boldsymbol v=
\boldsymbol k/k^0=0\,.$ Taking into account the transformational properties 
(9)--(12) we find 
planewave solutions of Eq.(13) in the final form\,:
\begin{equation}\nonumber
{\psi}_{+}(x)=\frac{\sqrt{k_0^2\!-\!{\boldsymbol k}^2}}{m(1\!-\!r)}
\left(\!\begin{array}{lll} \!\sqrt{k_0\!+\!\sqrt{k_0^2\!-\!{\boldsymbol k}^2}}\,\,\,\varphi \\
\phantom{a} \\
 \sqrt{k_0\!-\!\sqrt{k_0^2\!-\!{\boldsymbol k}^2}}\,\,\,(\boldsymbol n\boldsymbol\sigma )\varphi 
 \end{array} \!\right)e^{-i\,p_a\,x^a}\, ,
\end{equation}
\begin{equation}\nonumber
{\psi}_{-}(x)=\frac{\sqrt{k_0^2\!-\!{\boldsymbol k}^2}}{m(1\!-\!r)}
\left(\!\begin{array}{lll} \!\sqrt{k_0\!-\!\sqrt{k_0^2\!-\!{\boldsymbol k}^2}}\,\,\,
(\boldsymbol n\boldsymbol\sigma )\chi \\
\phantom{a} \\
 \sqrt{k_0\!+\!\sqrt{k_0^2\!-\!{\boldsymbol k}^2}}\,\,\,\chi \end{array} 
\!\right)e^{i\,p_a\,x^a}\, ,
\end{equation}
where the unit vector $\,\boldsymbol n\,$ indicates the direction of $\,\boldsymbol k\,,$ 
in which case $\,k^a\,$ and $\,p^a\,$ are related by (6). The bispinor 
density fields $\,\,{\psi}_{\pm}\,\,$ are normalized with the help of the invariant conditions\,\,:
$\,[({\nu_n {\bar{\psi}}_{\pm}\gamma^n{\psi}_{\pm}}/{{\bar{\psi}}_{\pm}{\psi}_{\pm}})^2]^{-3r/2}
{\bar{\psi}}_{\pm}{\psi}_{\pm}=\pm \,2m(1\!-\!r)\,.$ As for the dispersion relation 
(18), in terms of $\,k^a\,$ it takes the form 
$\,\sqrt{k_ak^a}=\pm\,m(1-r)\left[({\nu}_ik^i)^2/k_jk^j\right]^{r/2}\,.$ One of its solutions 
corresponds to massive fermions and, according to (5), admits the 
parametric representation 
by means of 3-velocities $\,\boldsymbol v\,.$ Another solution corresponds to massless fermions and 
has the form $\,k^a \propto {\nu}^a\,.$ Note at last that, in the limit $r=1,$ 
Eq. (13) takes the form 
$\,i\gamma^a{\partial}_a\psi\!-\!m{\nu}_a{\gamma}^a\psi =0\,$ and, after the local gauge 
transformation $\,\psi\to exp{(-im{\nu}_ax^a)}\,\psi\,\,,$ reduces to the massless Dirac equation. 

Summing up the results of the present work, we would like to emphasize that the 
spontaneous breaking of Lorentz symmetry does not necessarily signify the breaking of 
relativistic symmetry and may turn out to be a secondary effect induced by the spontaneous 
breaking of gauge symmetry. Here, the 10-parameter Poincar\'e group of an initial massless 
gauge-invariant theory is reduced to the 8-parameter inhomogeneous group of the generalized 
Lorentz transformations, which assumes in this case the role of the relativistic symmetry 
group of the corresponding vacuum solution of the theory. And vacuum itself, if it is regarded 
as space-time filled with a fermion-antifermion condensate, assumes anisotropic Finslerian 
geometry instead of Minkowski geometry. Within the framework of this picture the rearrangement 
of initial vacuum and the appearance of masses in the initial massless fields are not due to 
the standard Higgs mechanism but result from collective quantum effects peculiar to nonlinear 
dynamic systems. 

Reverting to the translationally invariant generalized Dirac equation 
(13), which was obtained from the requirement of the generalized 
Lorentz symmetry, we see that it is essentially nonlinear. However this nonlinearity 
disappears in two cases\,: firstly, if the anisotropy field, constant over the whole space 
(\,and more exactly, its magnitude $r\,$) tends to zero (\,in this case 
(13) changes to the standard massive Dirac equation\,), and 
secondly, if $\,r\,$ tends to its maximally attainable value equal to unity. In the latter 
case the anisotropy field turns out to be purely gauge while the massive 
fermion-antifermion field proves itself as the corresponding massless one. This means that 
the equation (13) describes the dynamics of the massive 
fermion-antifermion field in an 
anisotropic medium (in a relativistically invariant anisotropic condensate), in which case 
the effective inertial mass of the fermion-antifermion field has the dynamic origin and 
depends on the degree of order of the condensate, which should be a function of temperature.

Concluding the discussion of the nonlinear generalization of the Dirac equation, we note in 
addition that so far we have succeeded in constructing only simple, namely, planewave 
solutions of this equation. However, efficient algebraic-theoretical methods of constructing 
exact solutions for a wide class of nonlinear spinor equations have already been developed 
[15]. Using these methods, one can in principle obtain, also, 
other and, which is especially important, ``nongenerable'' families of exact solutions of the 
equation (13). As for the general conceptual problems 
relating to nonlinear generalizations 
of the Dirac equation [16], we hope to give more attention to 
them in our subsequent publications.

\subsection*{Acknowledgements}

We would like to thank Prof. H.-D. Doebner, Dr. V.M. Boyko and Dr. V.I. Lahno for useful discussions.
G.\,Yu.\,B. is also grateful to the Institute for Theoretical Physics  (\,G\"ottingen\,),
where this work was initiated, and to the Institute  of Mathematics (\,Kiev\,), where 
it was reported, for their  hospitality.

\newpage  

\begin{figure}[hbt]
\begin{center}
\epsfig{figure=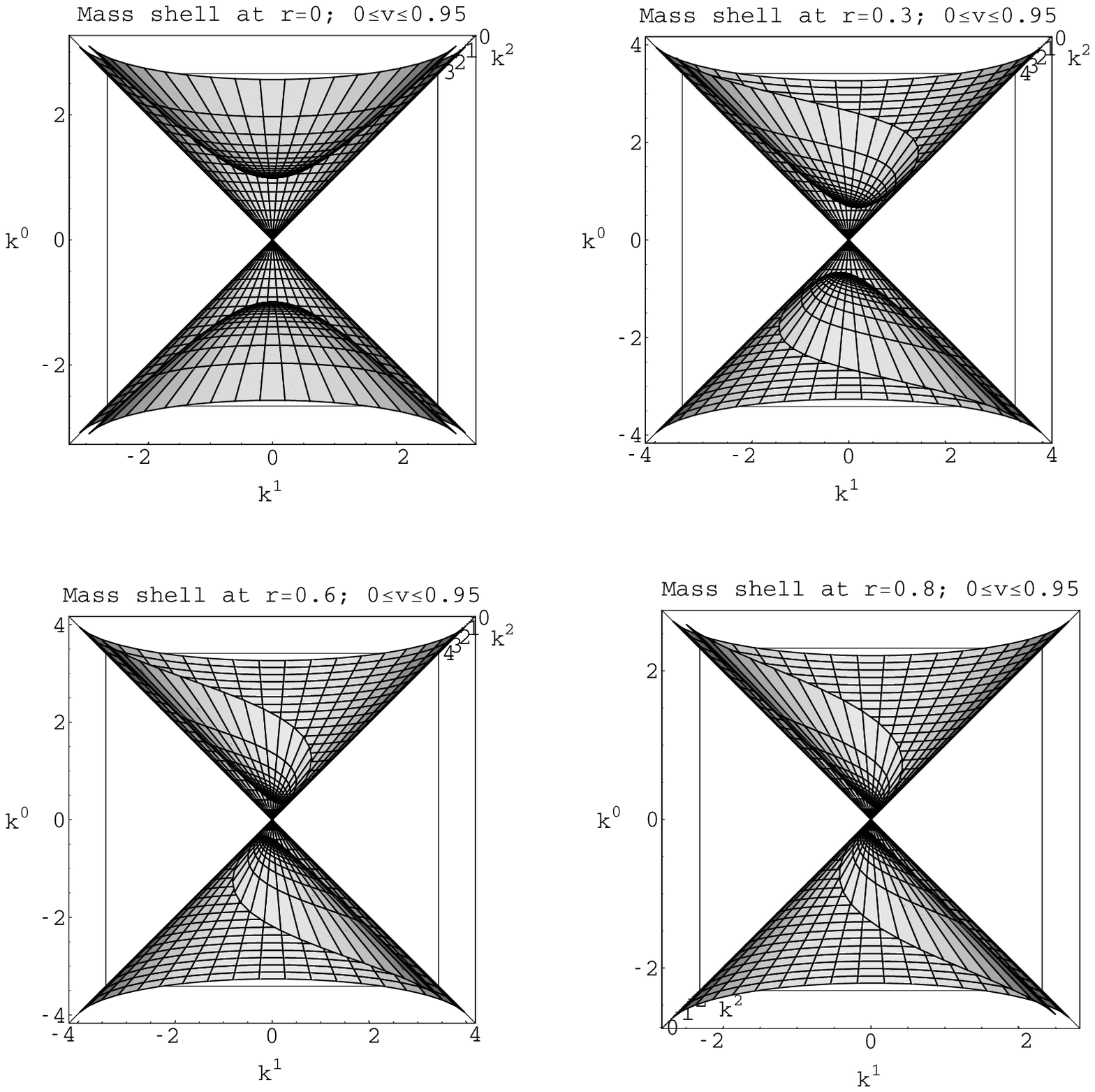,width=14.0cm,height=14.0cm}
\end{center}
\caption{Parametric 3D plots of the mass shells in a space of kinematic 4-momenta.}
\end{figure}

\end{document}